# Challenging Practical Features of Bitcoin by the Main Altcoins

Andy Spurr[1] and Marcel Ausloos[1,2,3]

[1]School of Business, University of Leicester, Brookfield, Leicester LE2 1RQ, UK
[2]Department of Statistics and Econometrics, Bucharest University of Economic Studies, 6 Piata Romana, 1st district, 010374 Bucharest, Romania
[3] GRAPES, Sart Tilman, B-4031 Liege, Belgium

**Abstract**

We study the fundamental differences that separate: Litecoin; Bitcoin Gold; Bitcoin Cash; Ethereum; and Zcash from Bitcoin, and draw analysis to how these features are appreciated by the market, to ultimately make an inference as to how future successful cryptocurrencies may behave. We use Google Trend data, as well as price, volume and market capitalization data sourced from *coinmarketcap.com* to support this analysis.

We find that Litecoin's shorter block times offer benefits in commerce, but drawbacks in the mining process through orphaned blocks. Zcash holds a niche use for anonymous transactions, benefitting areas of the world lacking in economic freedom. Bitcoin Cash suffers from centralization in the mining process, while the greater decentralization of Bitcoin Gold has generally left it to stagnate. Ether's greater functionality offers the greatest threat to Bitcoin's dominance in the market.

A coin that incorporates several of these features can be technically better than Bitcoin, but the first-to-market advantage of Bitcoin should keep its dominant position in the market.

## 1   Introduction

Much of the cryptocurrency market is focused on the supremacy of coins themselves, with less focus on the technical aspects that are the underlying factors for this success. Our motivation for this paper is to focus on these technical aspects, since the development of new coins, and improvements on existing coins benefit from this type of research.

Using "scholarly" methodology, we attempt to gauge how the market values such features. We thereby perform a qualitative analysis for a "better cryptocurrency choice" - although the final choice depends on the user's own criteria, of course. Concluding this paper, we make predictions as to which existing features are likely to be successful as this market continues to develop.

The cryptocurrency market has exploded in recent years, becoming an industry worth USD 470 bn (Coinmarketcap.com, 2018). Bitcoin (BTC) was the first currency of its kind; other alternative cryptocurrencies, 'altcoins', have since been created in an attempt to improve upon the way that their transaction value is digitally sent and received.

The key innovation that can be attributed to Bitcoin's rapid success is the use of the blockchain technology, but this is challenged (Chatterjee et al., 2018).

The blockchain forms a competitive environment by which users known as 'miners' validate transactions in order to receive a reward – the issuance of new coins. Miners commit their computing power to the network in an attempt to crack a complex mathematical puzzle, that once solved, allows a block of transactions to occur (Nakamoto, 2008). This is known as *proof of work*. The use of this puzzle prevents a single user from consistently being able to create blocks on the blockchain, preventing them from acting maliciously, by validating their own false transaction. On the contrary, miners are incentivized to remain honest, as these honest miners are eligible for the reward of coins. This distributed system of verification allows for a scenario whereby all users can form a consensus of trust, eliminating the need for a central authority to provide the role of overseer.

A user that wishes to cheat the system by creating a faulty blockchain would have to have access to more than 50% of the network's total computing power - which is a staggering amount (Hruska, 2017). Since coins are introduced by the system itself, they can be traced across accounts to their inception, allowing all users to trust the integrity of a given transaction. The use of a block prevents these accounts from performing double-spends, since all transactions within a block are checked to be simultaneously valid. A Bitcoin address consists of a public key (shown on the blockchain) and a corresponding private key (Bitzuma.com, 2018). Public key cryptography is used to combine these two values in order to create an unforgeable message signature. These keys are linked through a signature algorithm, which is known by the network.

The BTC network is pseudonymous, meaning that the addresses of wallets (public keys) are visible to all, since they are noted on the blockchain, which is public knowledge. The value of each transaction is known, but their respective owners are hidden. This has led many to question the morality of the network, since users can perform illicit activities under the protection of the anonymity that the network provides (Chen, 2016).

The network currently supports a system by which transacting users can offer a transaction fee in order to increase the likelihood that their transactions are included into an upcoming block. This transaction fee is rewarded to the miner that cracks the block's puzzle, on top of the newly created coins. Currently, transaction fees only account for 1% of mining revenue (Blockchain.info, 2018). Once the final Bitcoins have been issued, miners are likely to compete for transaction fees, as opposed to newly created coins.

One should notice that the Bitcoin protocol is merely a set of mathematical rules and code, which does not offer an individual anymore control than needed for an "average user". Nevertheless, this offers a problem to the flexibility of the network to change; this is overcome through a process known as a 'fork'. A fork is basically a software change, by which users change their protocol so that it matches that of other users. This makes changes to the Bitcoin network a matter of "democratic vote", whence is purely optional.

In Sect. 2, we provide a literature review on a reduced scale. Much information is, one should say *obviously per se*, in such an electronic world, obtained from the web. One could also say that most of the literature pertains to papers delivered at conferences, thus through conference proceedings. Remaining at a scholar level, we do our best to refer to papers which have undergone peer to peer review procedures. The research questions are next outlined in Sect. 3, before the methodology in Sect. 4. The "data" gathering is explained in Sect. 5, presenting the distinctive features of these six cryptocurrencies. A section, Sect. 6, on results with some discussion on qualitative and quantitative aspects follows. A conclusion section, Sect. 7, ends our report.

## 2  Literature review

Bohme et al. (2015) argue that Bitcoin is not completely decentralized, due to certain barriers-to-entry within the necessary functions of the network. Bohme et al. (2015) find that exchanges are not decentralized, since they are subject to legislation from the nation in which they operate, which effectively prevents a free market for this service. Exchanges in the USA must register with the Financial Crimes Enforcement Network, which require the payment of a license fee that can be up to six figures in value, whilst exchanges in Germany are required to have a minimum capital requirement of EUR 5 million. These significant capital requirements prevent the average person from setting up an exchange, leading to imperfect competition within this sector.

de Filippi and Loveluck (2016) consider the governance structure within the Bitcoin network and find that it is not completely decentralized. A question still raised by Chohan (2019). Recently, Manavi et al. (2020) proved that the "myth", for the Bitcoin-based networks, i.e. decentralization is only true in specific regions, i.e., it is not distributed homogeneously. In fact, Shaw and Hill (2014) had suggested that the governance of Bitcoin relies on leaders, forming oligarchic organizations, thus "centralized".

Hruska (2014) questions the decentralization of the Bitcoin mining process, as mining 'pools' have formed by which users pool their computing power in order to distribute mining rewards evenly across users, allowing for more consistent income from the mining process. Some of these pools offer this service in exchange for voting rights, which can allow these pools to have heavy influences on decisions such as voting on forks. This acts to centralize the network, since it increases the influence that a single voter might have on decisions. Eyal et al. (2016) study the impact of changing the block size and block times, and find that, although changes to this can increase the scale by which the network can handle transactions, these changes come at the cost of greater centralization. They find that fairness suffers, as any increase in the rate at which work is done benefits those with the capital to respond to such changes, while restricting users that cannot meet the increased demands in storage and processing speed.

Barber et al. (2012) find that the capped supply of Bitcoin will likely lead to an interesting scenario by which any appreciation in the currency's value will translate directly into an appreciation of the coin's price, as opposed to fiat currencies that are continuously printed, giving them an outlet for this growth. The main drivers of the Bitcoin price can be studied in many ways, as through a "Wavelet Coherence Analysis' (Kristoufek, 2015). Ammous (2018) studies the trends in currency printing, and finds that over the next 25 years, the supply of USD will likely increase by 372%, with GBP supply increasing even greater, at 530%. When compared to Bitcoin's 27% increase in the same period, the use of BTC as a store of value may be extremely appealing (Ammous, 2018), as it does not suffer the same loss in purchasing power as other fiat currencies. Indeed, BTC has an increased benefit of having a capped supply, whereas fiat currencies will likely continue to be created. This could threaten the purchasing power of fiat currencies, according to Barber et al. (2012), and therefore the strength of the governments behind them - whence political governance in the usual "old way". Barber et al. (2012) had suggested that this appreciation in value leads to a situation of hoarding, as users will wish to hold their coins due to the apparent future increase in value. Yet, Pieters and Vivanco (2017) found inconsistencies in altcoins market values.

Ammous (2018) predicts that any changes to the amount of coins issued to miners would be detrimental to the network, - since it would hurt a given coin's purchasing power, whilst affecting the predictability and credibility of the current supply system. He also suggests that a cryptocurrency is unlikely to be used as a unit of account until it becomes the most-used currency globally; until then, established currencies such as the US Dollar will maintain this primacy status. Yet, Gandal and Halaburda (2016) find that the network effect, when the value of a network increases in proportion to the number of its users, is strong for Bitcoin. Thus, Gandal and Halaburda (2016) suggest that newer currencies are coming too late if trying to establish themselves as the number one cryptocurrency in the ecosphere. For their reasoning, Gandal and Halaburda (2016) use Google trend data, as we do also below.

A blockchain analysis is an unavoidable feature of Bitcoin in its current form; the statistical methods at hand to law enforcement are largely kept secret. Garzik, a member of the Bitcoin development team is quoted by Madrigal (2017) as saying "It would be unwise to attempt major illicit transactions with Bitcoin, given existing statistical analysis techniques deployed in the field by law enforcement" during early development of the network. Yet, the extent to which Bitcoin has been used for illicit activity is significant (Trautman, 2014; Turpin, 2014). Between 2011 and 2013, the marketplace had handled 1,200,000 Bitcoin transactions, accounting for USD 1.2 bn at the time. The strength of the anonymity that Bitcoin provides was questioned by Reid and Harrigan (2013), analysing public key data on the blockchain, the authors find "patterns".

The reasoning behind the creator of Bitcoin choosing to remain anonymous has led some to suspect that Bitcoin is a Ponzi scheme or other type of scam, whereby a price bubble could form around the currency, with little intrinsic value other than speculation maintaining this value. Vasek and Moore (2018) rebut these claims, but do not rule out the potential given historical events such as the Dotcom bubble in the 1990s. Grinberg (2012) suggests that the anonymity that Bitcoin provides is not useful to most users, and that users will prefer the use of traditional commerce since it offers features that Bitcoin cannot. The availability of reversible payments for fraud protection and the familiarity that users have with pricing based in fiat currencies are key features for Grinberg that will prevent Bitcoin from being competitive in the commerce area. Grinberg (2012) goes on to say that Bitcoin is unlikely to thrive for online payments, and is more suited towards virtual payments such as in closed economies within video games. The growth of Bitcoin and other cryptocurrencies since this article suggest that Grinberg's predictions were not accurate.

Contrary to Grinberg (2012), Herrera-Joancomarti (2014) suggests that anonymity is one of the key properties of Bitcoin that has led to its large success. The author investigates the degree to which Bitcoin is anonymous by using information on the blockchain, as well as external information such as Twitter posts and forums to cluster users, and then attempts to identify them. Using this method, Herrera-Joancomarti suggests that 40% of Bitcoin users can be identified, given a certain level of available external information about them. Maurer (2016) investigates the anonymity of some cryptocurrencies that attempt to improve upon Bitcoin in this feature; as well as services that attempt to improve upon the privacy of Bitcoin itself. He finds that services that operate on top of Bitcoin are susceptible to theft, but are easier to implement than other options.

Concerning BTC as a currency which can be exchanged, like a fiat currency[1], thus with a fluctuating value, Sapuric et al. (2017) compare the daily and weekly returns, as well as the volatility of different digital currencies. Sapuric et al. (2017) find that the correlations between these currencies are low, suggesting that these could present good diversification opportunities for investing. Moreover, Sapuric et al. (2017) find that certain trends suggest that a bubble may be forming around cryptocurrencies, and that speculation is playing a key role in the industry.

Wang and Vergne (2017) compare five major cryptocurrencies. They use data on weekly returns, as well as details on developer activity and public interest in order to evaluate the strengths of these coins. They suggest that the development team behind a digital coin is key for maintaining value, and improvements in software are crucial when trying to maintain an initial technological advantage. Sovbetov (2018) finds that the key internal factors that contribute to a cryptocurrency's price are total market capitalisation, trading volume and volatility; while the key external influences are attractiveness, stock market movements and the price of gold.

Cerqueti el al. (2020) recently proposed a comprehensive study about the cryptocurrency market, evaluating the forecasting performance (from exchange rates with USD point of view) for three cryptocurrencies in terms of market capitalization. Urquhart (2018) investigates correlations between Google trend data for common search terms with Bitcoin's price to analyse investor attention.

Qin, Su and Tao (2020) compare Bitcoin's price and volume with global economic policy uncertainty (GEPU); they find that Bitcoin can generally be seen as a hedge, although some correlation exists between the two. Drozdz et al. (2019) support such a finding: they find that both Bitcoin and Ether demonstrate the statistical hallmarks, normally observed in ''mature markets'', like stocks, commodities, or the Forex. Drozdz et al. (2019) see the gradual emergence of a new and partially independent market in which not only Bitcoin but also the whole emerging crypto-market may eventually offer a hedge for fiat, gold and commodities.

---

[1] The technology and economic determinants of cryptocurrency exchange rates, in the case of Bitcoin, has been interestingly studied by Li & Wang (2017).

Su, Qin, Tao and Zhang (2020) compare Bitcoin's price with that of gold's. They find that Bitcoin undermines the hedging ability of gold, but not so much as to threaten either's utility within a portfolio. Li, Tao, Su and Lobonţ (2018) find bubble components in Bitcoin's price at several points, including between 22nd December 2016 and 5th January 2017.

# 3 Research Questions

Focusing on BTC, the above literature review identifies a few key areas of concern for all cryptocurrencies, namely: privacy; centralization; scalability; supply; and intrinsic value. Thus, we have decided to compare factual information on several top coins in the market, in order to evaluate the success of features that differ from those implemented in the original Bitcoin protocol. The goal is to isolate specific features, use different sources of data in order to evaluate how the community has responded to such fundamental changes, and propose an objective decision support set of criteria. This evaluation of features seems to be important when predicting the chosen features of new coins in the market, and furthermore what features the most successful coins of the future might possess (Chatterjee et al., 2018). This leads us to propose the following research questions:

How does the cryptocurrency community value the feature of

- Greater anonymity?

- Decreased block times?

- Increased block size?

- Application Specific Integrated Circuit resistance?

- Level of intrinsic value or of (better) functionality?

and, last but not least,

- What features can be expected to be found in future cryptocurrency projects for optimizing interesting deviations from BTC?

## 4 Methodology

In order to reach qualitative and quantitative answers to the above research questions, we compare the total market capitalization, price and volumes of these coins across a period of eight months, from 11/08/2017 to 1/04/2018, and pursuit their relative positions in order to track community adoption. This eight-month period coincided with a steep rally for the market as a whole, and volume issues for Bitcoin (Begušić et al., 2018) created an interesting opportunity, as it encouraged users to experiment with alternate coins. On Fig 1, we emphasize some of the key world events occurring during this time interval. The introduction of futures contracts for Bitcoin was likely the main catalyst for the increase in price, and corresponding liquidity issues (Shen, 2017). Subsequent gains in this period would suggest that the community valued a coin, whilst losses would suggest that other options were preferred.

Correlations between Google Trend data and these metrics further support our investigation, - since a relationship between research and adoption is an indicator of how technical features of a cryptocurrency are desired by the market.

In the following, we compare Bitcoin and five key altcoins in order to assess their respective features. By choosing coins with a high market share, we hopefully eliminate the possibility of scams and schemes, since a high market share represents a high level of trust across the community. Top ranked coins are more likely to represent the most successful features currently in the market, and are therefore more likely to have these features included in the design of future coins.

Choosing coins that differ from Bitcoin in as few ways as possible allowed for a high level of comparability, yielding more useful results when isolating specific features implemented by these currencies.

We include Litecoin in the study, since it was the second cryptocurrency launched in the market, and has maintained a strong position in the market ever since. This second-to-market advantage could help to reduce the impact of qualitative features that Bitcoin possesses such as its strong brand name and notoriety, since Litecoin should possess similar characteristics, albeit on a weaker scale.  Litecoin was also a useful pick for the study, since its resistance to ASIC processors was overcome, allowing for analysis on a how a once useful feature was eliminated by technological advances. This differs from the other features in the study, as price data could offer insights into how valuable a feature is through its loss in functionality, rather than how the community has reacted to its launch.

Forked coins, as defined here above, proved to be invaluable, since they are heavily based on Bitcoin, being that they exist in the form of updates to the network. Both Bitcoin Cash and Bitcoin Gold differ in one key area from Bitcoin, which should allow some pertinent finding whence a strong isolation of how these features are welcomed by the market.

We do not consider Ripple in this study, although it was ranked in the top five for market share at the time of our choosing this investigation, as Ripple does not use distributed ledger technology as seen in the other cryptocurrencies on the market. Notice that Ripple represents a different genre of currency, since it is highly controlled by governing bodies.

Several top ranked coins offer privacy improvements to Bitcoin; we chose to study Zcash in particular, as it is more closely related to Bitcoin. We could not find a suitable coin that offered perpetual supply, one which was sufficiently similar enough to Bitcoin to allow for strong comparability. We chose to analyse Ether in this study, as although it offers differences to Bitcoin in many significant areas, it has performed exceedingly well in the market, and offers improvements in functionality and intrinsic value, which has been a key critique of Bitcoin in the past. Although comparability is relatively low in respect to this, we consider the fundamental changes that Ether offers to be significant enough to include it in the study, since data here should be crucial in answering research questions on what to expect in cryptocurrencies in the future.

# 5 Data

We sourced data on price, transaction volume and market cap over time using coin-marketcap.com. We chose to use the daily 'high' prices, as opposed to open and closing values, since we wanted some data to emphasize the potential volatility of the currencies, as large swings can occur in a 24-hour period. Outliers have often some significance, indeed.

In Fig. 2, we display such daily 'high' prices of each coin type over the course of a year, after normalizing them so that on the first day, each price is equal to '1'; each other data point is presented as a ratio of this figure,

$$R(t) = Price(t)/Price(0) \qquad (1)$$

The basic statistical characteristics of the raw data are provided in Table 1 (top); those of the ''normalized data'' are found at the bottom of Table 1. The skewness and kurtosis are of course identical.

|  | min | Max | mean | median | RMS | StdDev | StdErr | Skewn | Kurtosis |
|---|---|---|---|---|---|---|---|---|---|
| Bitcoin | 3664.8 | 20089 | 9087.3 | 8468.7 | 10046 | 4291.6 | 280.55 | 0.6461 | -0.5157 |
| Litecoin | 44.830 | 375.29 | 139.89 | 104.71 | 165.12 | 87.927 | 5.7480 | 0.6640 | -0.7048 |
| BitcoinCash | 306.52 | 4355.6 | 1246.0 | 1133.1 | 1493.1 | 824.50 | 53.899 | 1.0799 | 0.7858 |
| Zcash | 177.40 | 955.27 | 361.54 | 306.31 | 392.60 | 153.37 | 10.026 | 1.1988 | 0.9628 |
| Ether | 257.00 | 1432.9 | 598.38 | 473.17 | 677.53 | 318.47 | 20.819 | 0.7601 | -0.6145 |
| Bitcoin | 0.9960 | 5.4594 | 2.3015 | 2.7300 | 1.1663 | 1.3602 | 0.07624 | 0.6461 | -0.5157 |
| Litecoin | 0.9472 | 7.9292 | 2.2124 | 3.4888 | 1.8577 | 3.4512 | 0.12144 | 0.6640 | -0.7048 |
| BitcoinCash | 0.8729 | 12.404 | 3.2267 | 4.2519 | 2.3479 | 5.5128 | 0.15349 | 1.0799 | 0.7858 |
| Zcash | 0.7556 | 4.0686 | 1.3046 | 1.6721 | 0.6532 | 0.4267 | 0.04270 | 1.1988 | 0.9628 |
| Ether | 0.8312 | 4.6340 | 1.5303 | 2.1912 | 1.0300 | 1.0608 | 0.06733 | 0.7601 | -0.6145 |

Table 1. Table of usual statistical characteristics for (top) raw data, (bottom) ''log-normalized'' data, respectively.

We sourced data on Bitcoin transaction fees from Quandl.com (2018), and took charts of the market cap and hash rate of Litecoin from Bitinfocharts.com. We sourced all data on prices in US Dollars, in order to remove any fluctuations in exchange rates from our results. We took worldwide search data from Google Trends (2018). This data could have been more useful, as Google chooses to provide relative figures on searches, as opposed to actual values of searches. The values are pre-rounded, meaning that low relative search volumes appear as 0, when there are likely search volumes in this period, albeit small values. To combat this issue, we only used this data when analysing trends, in order to mitigate these inaccuracies.

## 5.1 Data Selection

Notice that we chose not to evaluate Bitcoin Gold within (Fig. 2 and Table 1), since its introduction to the market in October 2017 did not present sufficient data points with which to track the coins for a meaningful period of time. Instead, we chose to begin the tracking of these prices 20 days after Bitcoin Cash had entered the market; this period represented large price movements, largely due to the uncertainty of the community around this time. For instance, the price of Bitcoin Cash varied from USD 756 per coin, to as little as USD 223.70 within these first 20 days. The price after this period was much more stable, which allowed for more relevant analysis on price movements.

We feel that comparisons over a period of eight months, without any further argument, is sufficient, as this allows for a clear analysis of the trends between coins, accounting for short term-variation. The price and market cap of cryptocurrencies tend to fluctuate as relevant news is transferred across the market, meaning that at any given point in time, their relative values may not represent long-term trends, as significant news may affect one coin, but not another. Data taken beyond this eight-month period would not be as insightful for this particular study, as both Bitcoin Cash and Bitcoin Gold had not formed before this time, and 2017 saw a significant increase in the production of new coins, largely through the initial coin offering (ICO) market. This presence of new coins allowed for a wider range of intra-coin comparisons; moreover, any coins that established success during this period have faced more intense competition, thereby suggesting that their features are of a higher quality.

## 5.2 Altcoin Selection and their Main Features

Many coins on the market have made changes to the initial Bitcoin blueprint in order to meet gaps in the market or to create a cryptocurrency that is superior. With this in mind, we have carefully selected coins on the market which we believe to be genuine in their intentions – to improve on Bitcoin in order to establish themselves as the leading cryptocurrency. We have noticed that Ong et al. (2015) have compared the top 10 currencies, at their time of writing, in order to evaluate their relative potential. Ong et al. (2015) used data from Reddit and Twitter in order to identify the popularity of each coin. Not disregarding at all the report by Ong et al. (2015), we dare to claim that our level of investigation is more profound, – but, we admit it, over six currencies only.

### 5.2.1 Litecoin (LTC)

Litecoin was launched by Charlie Lee, a former Google employee, on October $7^{th}$, 2011, in an attempt to make slight qualitative changes to the original Bitcoin protocol. Litecoin is heavily based on Bitcoin, but varies in a few fundamental ways, as we point out next.

Firstly, Litecoin shortened the time with which a block is created from the original 10 minutes to 2.5 minutes. The idea was that by reducing the time taken to add blocks to the chain, more transactions could be completed across the network in a given period of time, and with greater speed. This is extremely useful for merchants, since it decreases the time that they must wait before being sure that a transaction has been validated by the network. This however comes at the cost of orphaned blocks, which are those that have been solved by miners, but not chosen as the continued path along the blockchain. This leads to a situation by which miners have solved the puzzle, but do not receive a reward, effectively wasting their time and energy. As block times decrease, the risk of orphaned blocks increases exponentially (Rosic, 2017b).

In December 2017, Litecoin suffered scalability issues, whereby miners were unable to process the sheer numbers of transactions at a sufficient rate. The solution that Litecoin found to this problem is known as 'Segregated Witness' or 'SegWit'. This effectively shortens the amount of details included in a transaction, allowing for more transactions to fit into a block. Bitcoin has adopted the same solution for the scalability issues that it suffered in the same period; both networks have begun experimenting with off-chain solutions through technology known as the 'Lightning Network'. This technology allows for transactions to occur outside of the blockchain, which are then netted against each other, and finally settled on the chain. This hopes to decrease the congestion within both networks, allowing for quicker transaction verification and lower fees.

Litecoin chose to use a different algorithm to that of Bitcoin which relies more on a mining computer's memory, as opposed to the original algorithm that offered better efficiency with greater processing power. The idea behind this was to offer greater centralization across the network, as it made it easier for regular users to compete for the prize against the hardcore miners that have access to greater capital, and therefore more specialised equipment. This was the situation until ASIC miners were developed for Litecoin that can efficiently overcome this restriction, reducing the value of this feature when compared against Bitcoin.

Finally, Litecoin differs itself from Bitcoin by increasing the total supply of coins from 21 million to 84 million. This offers merely a psychological advantage when compared to Bitcoin, since both coins are divisible by up to 8 decimal places, and the value of each coin is its proportion of the currency's total value. This creates a situation where an identical trade would cost 2 units in Litecoin as opposed to 0.5 units in Bitcoin, - which may be more practical in appearance, and beneficial towards those with limited numerical skills.

Thus, the key difference with Bitcoin is: "Decreased block times".

### 5.2.2 Bitcoin Cash (BCH)

Bitcoin cash was created in a fork (a software upgrade) from the original Bitcoin blockchain on the 1st August 2017, and was intended to solve Bitcoin's scaling problems. In July 2017, a majority of Bitcoin users voted in favour of the fork, and went forward with implementing SegWit in order to tackle the congestion on the network. Some users felt that implementing SegWit would unfairly benefit users that wished to treat Bitcoin as a digital holding asset, rather than a currency, and so opted to take a different approach. This approach was to increase the block size from 1MB to 8MB, allowing for the amount of transactions taking place within a 10-minute period to increase eightfold. This acted to decrease transaction costs, since there was less competition between users wishing to push their transactions onto the chain. The price at the time of the fork was incredibly volatile, before settling at an exchange rate of around 0.1 BTC per 1 BCH. Futures before the fork were offering up to 0.5 BTC for 1 BCH, emphasizing the price uncertainty around this time.

Bitcoin Cash is effectively an updated version of Bitcoin, meaning that anyone owning Bitcoin at the time of the fork was also entitled to the same number in Bitcoin Cash. For instance, at the date of the fork, a wallet that contained 10 BTC would now own 10 BCH as well, since their address now exists for the two different currencies. This address could then send 10 BTC to a Bitcoin address, and 10 BCH to a Bitcoin Cash address separately. This is therefore a split in the value of the Bitcoin currency at this point into two segments, while the user base is also divided between advocates and critics – those who use Bitcoin, and those that use Bitcoin Cash. Due to this split, some users worry about the splintering of the Bitcoin community every time a hard fork is proposed; indeed, the strength of a currency is the size of its user base, and a user's ability to transact with others using the same denomination. This is known as the "network effect".

An increased block size arguably helps to resolve the scaling issues that Bitcoin faces, but critics argue that the increased size leads to greater centralization, since larger blocks require greater work, and this concentrates the mining process to larger mining nodes (Bogart, 2017). Advocates of Bitcoin Cash see it as being more suitable as a medium for exchange, while Bitcoin is seen as being more suitable as a store of value. Bitcoin Cash plans to hold another fork in May 2018, with a proposed further increase in block size to 32MB. This is expected due to Bitcoin Cash's plans to experiment with Turing Completeness, which would likely give the currency the functionality for 'smart' contracts, such as that of the Ethereum network (Bogart, 2017).

Thus, the key difference against Bitcoin is: "Increased block size".

### 5.2.3 Bitcoin Gold (BTG)

Bitcoin Gold, like Bitcoin Cash, was formed due to a fork from the Bitcoin blockchain, on October 24th, 2017. Some members of the community felt that the invention of ASIC miners created a market monopoly by which users would have to purchase this expensive equipment in order to remain competitive in the mining process. This theoretically leads to greater centralization within the network, since there are greater barriers to entry in the mining process. Some argue that this goes against Satoshi's vision of a decentralized currency, and therefore the nature of Bitcoin as it was first envisioned.

The solution to this was to change the algorithm within Bitcoin from 'SHA-256', which is currently seen in Bitcoin and Bitcoin Cash, to 'Equihash', which as of April 2018, is resistant to ASIC technology. This allows users to mine Bitcoin Gold on a more equal playing field, as anyone can start mining Bitcoin Gold with a standard off-the-shelf laptop. A team of developers designed this change and included a scenario by which 100,000 coins were automatically mined during the fork process, which are intended to fund further development to the currency.

Thus, the key difference with Bitcoin is: "Application Specific Integrated Circuit resistance".

### 5.2.4 Zcash (ZEC)

Zcash was launched on the 28th October 2016 by the Zcash Electric Coin Company and differs from Bitcoin in a few key areas. Like Litecoin, Zcash opted to shorten its block times to 2.5 minutes, from the original 10 minutes proposed by Satoshi Nakamoto. Since this coin was released five years after Litecoin, the Zcash development team had the opportunity to compare both Bitcoin and Litecoin's success on this matter, and their decision to follow Litecoin may imply that 2.5 minutes block times are optimal.

Zcash elected to support its development team by transferring 10% of the coins mined to chosen groups involved in the currency, including employees and investors. This is not concerning for miners, since they calculate their profitability based on the value of the coins that they receive; not the value that is issued by the system for a given block. The existence of a development team here could present problems for some, since the decision making of the development team can have large implications for the coin's price. As Satoshi chose to remain anonymous, his influence is limited here, which may be favourable to some since there is a reduced threat of price movements based on the actions of a single person or group. The key fundamental innovation by the Zcash team uses zero-knowledge proofs in order to shield transaction values, increasing the privacy within trades (Zcash, 2018). Zcash takes the concept of "*pseudonymity*" (where addresses are visible, but owners are unknown) within Bitcoin to a scenario in which transactions are truly anonymous. This has been made optional to users, allowing them to choose whether to disclose certain details of their trading activities, which can be useful for circumstances such as declaring details for tax purposes. The use of these proofs has made Zcash the first truly fungible coin, meaning that each coin is interchangeable with another.

Zcash similarly changed its algorithm to one that places more emphasis on memory, known as 'Equihash'. This was likely done for the same reasons as Litecoin's change, but unlike Litecoin, there is yet to be an ASIC designed to overcome this. This makes the Zcash algorithm superior in this instance, and should provide Zcash with the opportunity to remain more decentralized than competing coins that have specially developed ASIC processors. Zcash also elected to implement a block size of 2MB, twice that of Bitcoin, in order to account for the increased data within its more complex transactions. This does not provide a significant benefit towards its ability to handle transaction volumes, as transaction data is larger due to the elements that establish anonymity.

Thus, the key difference versus Bitcoin is: "Increased anonymity".

### 5.2.5 Ether (ETH)

Ethereum, the network powered by the currency Ether, was released on 30th July 2015, and was met with difficulties immediately (Ethereum.org, 2018; En.wikipedia.org, 2018). Funds of 11.9m Ether had been generated in order to reward investors of the currency, but problems in the security of this process led to the raised coins being coded to transfer ownership to an unknown attacker. The funds amassed over USD 150 m at the time of the attack; the community was very divided on how to react to this situation. With limited time to reach a verdict, the solution of a soft fork that blacklisted the funds was rolled out, preventing the attacker from realising the profits of their scam. This essentially locked the funds in place, preventing anybody from being able to access them. After much debate in the community, a larger portion of the community decided to implement a hard fork that allowed the trapped funds to be returned to their original owners. This splits Ethereum into two sides: Ethereum classic (ETC) and Ethereum (ETH), with the latter being the post-fork blockchain (Rosic, 2017a). This immediately set a precedent for the Ethereum network, since a community-wide decision was made to tamper with existing code on the blockchain, which unsettled those that expected irreversibility of transactions at all times. Bitcoin has currently maintained its status of irreversible transactions, despite many large-scale cases of theft and scams.

Unlike the other coins mentioned in this study, Ethereum is significantly different to Bitcoin in terms of functionality. Rather than being based on Bitcoin as a blueprint for digital currency, Ethereum uses the concept of the blockchain in order to develop a decentralized Turing-complete virtual machine. This is effectively a decentralized computer than can process codes and scripts across the network. This allows for interesting concepts such as 'smart' contracts, which offer mathematically binding agreements that execute under given circumstances. The full potential of this technology has not been fully realised, but Ether is used as the system's native currency in order to pay fees in return for this functionality.

Ether can be traded between wallets, and is generated into the atmosphere by miners, similarly to that of Bitcoin. This gives it the same functionality of being a currency, with added benefits of being necessary for functions other than currency. The lack of presence of intrinsic value is often thought to be the biggest issue within Bitcoin; that has caused many to speculate that Bitcoin is an asset bubble: there is no secondary function to act as a safety net, nor offer guidance as to what value should be put onto it. The Ethereum network may be a solution to this issue. One could expect the price of Ether to be less volatile than that of other cryptocurrencies, since its price has more contributing factors than just investor speculation. The inner workings of Ethereum are significantly different to Bitcoin, such as it having 14 second block times, an unlimited supply, and its blocks are capped by their cost of computation. This makes comparisons between its performance and Bitcoin's fairly redundant when trying to isolate the value of specific changes. There is value in comparisons when identifying possible future success however, since Ether has consistently taken the spot for market share in the cryptocurrency atmosphere. We have chosen to include it in this study, because it is informative to do so when searching for features that may be present in the currency of the future. In this case, we use the intrinsic value that Ether has, beyond being a medium of exchange, as its key differing feature against Bitcoin - although many other differences are present.

Thus, the key difference of Ether vs. Bitcoin is: "Intrinsic value is better functionality".

## 5.3  Synopsis of distinctive features

In summary, distinctive features of these six cryptocurrencies are proposed in Table 2.

| Coin name | Bitcoin | Litecoin | Bitcoin Cash | Bitcoin Gold | Zcash | Ether |
|---|---|---|---|---|---|---|
| Unit | BTC | LTC | BCH | BTG | ZEC | ETH |
| Block size | 1MB | 1MB | 8MB | 1MB | 2MB | Varied |
| Block time | 10 min. | 2.5 min. | 10 min. | 10 min. | 2.5min | 10-19 seconds |
| ASIC resistance | No | No | No | Yes | Yes | No |
| Privacy | Pseudonym. | Pseudonym. | Pseudonym. | Pseudonym. | Anonym. | Pseudonym. |
| Supply | 21,000,000 | 84,000,000 | 21,000,000 | 21,000,000 | 21,000,000 | unlimited |
| Function | Currency | Currency | Currency | Currency | Currency | Token |
| Forked? | No | No | Yes | Yes | No | Yes |
| Key Innovation | - | Shorter | Larger blocks | ASIC resistance | Anonymity | Functionality |

Table 2: The table contains a summary of distinctive features of the examined cryptocurrencies

# 6  Results and discussion

Having sorted out the key features from the historical information content, we may come up to some synthesis for the main differences, i.e., using logical comparisons with regards to Bitcoin. First, let us focus our attention to Litecoin and Bitcoin: the key differing feature between Litecoin and Bitcoin is its 2.5 minutes block size, which is four times shorter than Bitcoin's.

This feature does not seem to be valued too highly by the community, as Litecoin's market share represents less than 10% of Bitcoin's. During the eight-month period (Fig. 2), Litecoin actually performed the best, with the greatest proportionate price increase across the period. This was surprising, as Litecoin is the second- oldest cryptocurrency both in the study, and in the market as a whole, which would suggest that any growth in this coin may have occurred before now. As the tested period contains a period of high congestion for Bitcoin (Bitinfocharts.com, 2018), one can assume that users looked at alternative cryptocurrencies in order to avoid paying the high transaction fees associated with this congestion. As Litecoin maintained its value more efficiently than the other tested currencies, this suggests that the users that explored this coin during the time of congestion were happy with the currency's functionality. Litecoin should have a reputable brand name because of its maturity in the market, which could suggest why users chose this as an alternative to Bitcoin. Litecoin's market capitalisation against Bitcoin's (Bitinfocharts.com, 2018) tripled from April 2017 to April 2018, which suggests that the coin is gaining momentum. This however remains to be quite insignificant, with a market capitalisation of only 6% by April 2018. This suggests that the currency's shorter block times are not too valuable to the community; the growth in the currency's market capitalisation against Bitcoin suggests a gaining popularity.

The removal of ASIC resistance, as shown by the increase in hash rate (Bitinfocharts.com, 2018) was met with an increase in market capitalisation (Bitinfocharts.com, 2018), which is somewhat surprising, since an intended feature of the currency was overcome. This suggests that the community does not greatly value this feature, but sees the decentralization of Bitcoin as being sufficient.

If all other variables were consistent, Litecoin would appear to be superior to Bitcoin because of the benefits of a shorter block time. If the currency was able to solve the problem of orphaned blocks, Litecoin would have a better functionality than Bitcoin, and may have the advantage of these benefits being introduced at the coin's initiation, since a hard-fork in the Bitcoin network to implement shorter block times would likely be considerably difficult, if not infeasible.

Next, let us compare Bitcoin Cash and Bitcoin. The key differing feature for Bitcoin Cash is its larger block size. This allows the network to handle greater transaction volumes, but comes at the cost of greater centralization. Bitcoin Cash saw the largest price increases across the eight-month period (Fig. 2), suggesting that many users opted to use Bitcoin Cash during Bitcoin's period of high transaction fees (Bitinfocharts.com, 2018). The price of Bitcoin Cash had the largest volatility in this period (Fig. 2; see also the mean and Standard Deviation of the log-normalized data in Table 1), which is not surprising given that the currency had been released 20 days prior. Bitcoin Cash did not hold its value as well once the congestion period settled, suggesting that users chose to return to Bitcoin once it had implemented SegWit to deal with this issue. This suggests that those users did not value the greater centralization of Bitcoin Cash, as its functionality is the same, if not superior to Bitcoin, given their similarities. This notion is supported by a low correlation between its market capitalisation and transaction volumes (Table 3), which was the lowest of the coins tested. A low correlation here suggests that the market capitalisation figure is heavily inflated, since the network effect dictates that the value of a currency should be proportionate to its users. One can suspect that a scenario exists in which a small group of large capital investors are manipulating the supply and demand of the currency in order to give it the appearance of a greater success than is actually the case, in order to profit from the increased price of the currency. This would be a logical process for a more centralized currency, since investors with larger capital stand to gain the most.

| Crypto-currency | Google searches vs. transaction volumes | Market capitalisation vs. transaction volumes |
|---|---|---|
| Bitcoin | 0.801 | 0.921 |
| Litecoin | 0.415 | 0.631 |
| Bitcoin Cash | 0.311 | 0.474 |
| Bitcoin Gold | 0.727 | 0.446 |
| Zcash | 0.569 | 0.739 |
| Ether | 0.411 | 0.832 |

Table 3: The table contains the values of the correlation coefficients between worldwide Google searches and transaction volumes from April 2017 till April 2018, and the corresponding correlations between market capitalisation and transaction volumes.

Bitcoin Cash also had the lowest correlation between Google searches and transaction volumes (see Table 3), which suggests that research into the currency did not transfer into more users of the currency. This indicates that these users did not consider Bitcoin Cash suitable or worth its risk when seeking a currency in order to carry out transactions.

Next, consider Bitcoin Gold. The key differing feature for Bitcoin Gold is its offer of greater potential decentralization due to its resistance to ASIC processors. This has not been met with the support of the community, since the value of Bitcoin Gold has steeply declined since its inception (Bitinfocharts.com, 2018). This is mirrored by the community's reaction to Litecoin's resistance being overcome, which actually caused the currency's value to rise, rather than fall, as would be the case if the feature was valued highly by the community. Bitcoin Gold has a weak correlation between its market capitalisation and transaction volumes (see Table 3) - which suggests that like Bitcoin Cash, its market capitalisation figure is inflated. This is not for the same reasons, however, as it appears that Bitcoin Gold's market capitalisation has not yet fell in line with its declining user-base. Thus, Bitcoin Gold's market capitalization will likely continue to fall, unless the currency implements a significant update that could reinvigorate its user-base.

Next, notice that the key differing feature between Zcash and Bitcoin is its increased privacy. Zero-knowledge proofs allow users to conceal both the identities of users, as well as the values that are being traded. This does not appear to hold significant value to the community, since Zcash's price actually fell across the eight-month period (Fig.2).

On the contrary, Fig. 3 suggests that the currency was mostly searched for in China, where tough regulations have been placed on cryptocurrencies (Hsu, 2018). This supports the idea that Zcash holds significant value to users that require its greater anonymity, such as users in restrictive countries, but the majority of users do not find this feature particularly meaningful. This is supported by the correlation between Zcash's Google searches and transaction volumes, as this figure is higher than Litecoin, Ether and Bitcoin Cash (Table 3). This shows that research into the currency was met with a higher rate of users choosing to use this currency, indicating that the feature of greater anonymity holds some niche value to a minority of users. The findings corroborate those of Manavi et al. (2020) concerning regional importance.

"Finally", Ether is significantly different from Bitcoin (Guo et al., 2019), but presents the greatest threat to its position of having the greatest market share. Ether managed to gain 80% of Bitcoin's market capitalisation in June 2017. This however fell significantly by December of the same year, indicating the presence of an exodus of users around this time. By March, the currency had made a revival, but not to the extent of June 2017. Ether's price was fairly stable during the eight-month period (Fig. 2), suggesting that the greater functionality may act to reduce the volatility of the currency, which could present an advantage for investors, but is unlikely to benefit users greatly, since they are likely to exchange their digital balances for a fiat currency.

Ether has a high correlation between its market capitalisation and transaction volumes (see Table 3), suggesting the presence of a healthy user-base. Its low correlation between Google searches and transaction volumes however suggests that this currency was not adopted greatly by those that undertook research (Table 3). This could be due to the fact that Ether is quite unlike the other currencies tested, due to the greater complexity of the Ethereum network. Users that simply wish to send value digitally would likely find a currency such as Bitcoin to be simpler, since the use of smart contracts and other functionality is not necessary to achieve this goal. We predict Bitcoin Cash to be the biggest rival to Ether, if its implementation of Turing-complete functionality proves to be successful.

We predict that Ether will likely remain the more successful network of the two, since Ethereum's decentralization is more akin to that of Bitcoin, than Bitcoin Cash, which appears to be a significant downfall for the latter.

# 7 Conclusions

The currencies that offer significant advantages are these same currencies, namely Bitcoin, Ether and Zcash. These currencies offer unique benefits to users that are fundamentally useful, while the other currencies tested in the research appear not to be divergent enough to be favourable against the strong popularity of Bitcoin. Their low correlations between market capitalisation and transaction volumes are indicative of posturing and superficial inflation, in an attempt by holders of these currencies to attract users, which would likely increase their price, thus offering profitability for these holders.

When looking to invest in up and coming coins, it is important to find those with features that cannot be implemented into existing cryptocurrencies through forks. The decentralized nature of cryptocurrencies creates a scenario by which any intellectual property can be capitalised upon, since there is not a central body that could face ramifications by doing this.

If Bitcoin could implement shorter block times, Litecoin would likely lose most of its value, but if it were not possible, Litecoin is holding significant potential in this regard. The wider community does not seem too interested in greater anonymity; thus, a market will exist for Zcash, but not to a large enough extent to threaten Bitcoin. Any updates to the network that threatens centralization will likely be neglected by the community, since cryptocurrencies offer the potential for economic freedom that has never been seen until now. Bitcoin Cash's implementation of Turing completeness (Wright, 2016) will be interesting, as it has wider implications for the success of the Bitcoin network's ability to do the same, which could threaten Ether's advantages of functionality. Bitcoin Gold's extension to centralization, along with Litecoin's attempt to do so has not been appreciated greatly by the community, suggesting that the level offered by Bitcoin is sufficient.

As ASICs have overcome Litecoin's attempts to resist, similarly, Zcash will likely have ASICs developed to overcome its resistance; the profitability for doing so is a tempting offer for developers. If this was to happen, Zcash's stock will be rising, in a similar fashion to Litecoin's. If future developed coins were able to threaten Bitcoin's dominance, then these coins will have to overcome the problem of scalability permanently. Future coins would do well to find an alternative system to hard forks, in order to reduce the splintering of the currency's value on each implementation of an upgrade. Greater privacy is not likely to be included here, nor is greater levels of centralization, but any decrease in such a feature will decrease a given coin's ability to remain competitive. A system with greater intrinsic value may be beneficial to investors, and in the long-term users, since large fluctuations can hurt the spending power of a currency.

Cryptocurrencies are born in the internet age, and as such do not exist in an economic vacuum, free from online interference. Social networks play a huge part in the spread of information, both positive and negative, as well as true or false, whence are expected to play an active role on currency markets (Dhesi and Ausloos, 2016). Here lies an exciting opportunity to take publicly available correspondences, and engagement metrics and connect them to key performance data in order to map the effects of social influence.

If this were to be studied further, we would recommend data from social media outlets, such as Twitter, Facebook and Reddit, to be gathered and analysed supplementary to our paper in order to create a picture of how any news of coins would be to attracting users of a given coin. Within this, one should be interested to see how a coin with an uncapped supply is met by the market, since that is still a key issue amongst the Bitcoin literature.

Figure 1: Annotated events, December 2017 – January 2018 Bitcoin rally.

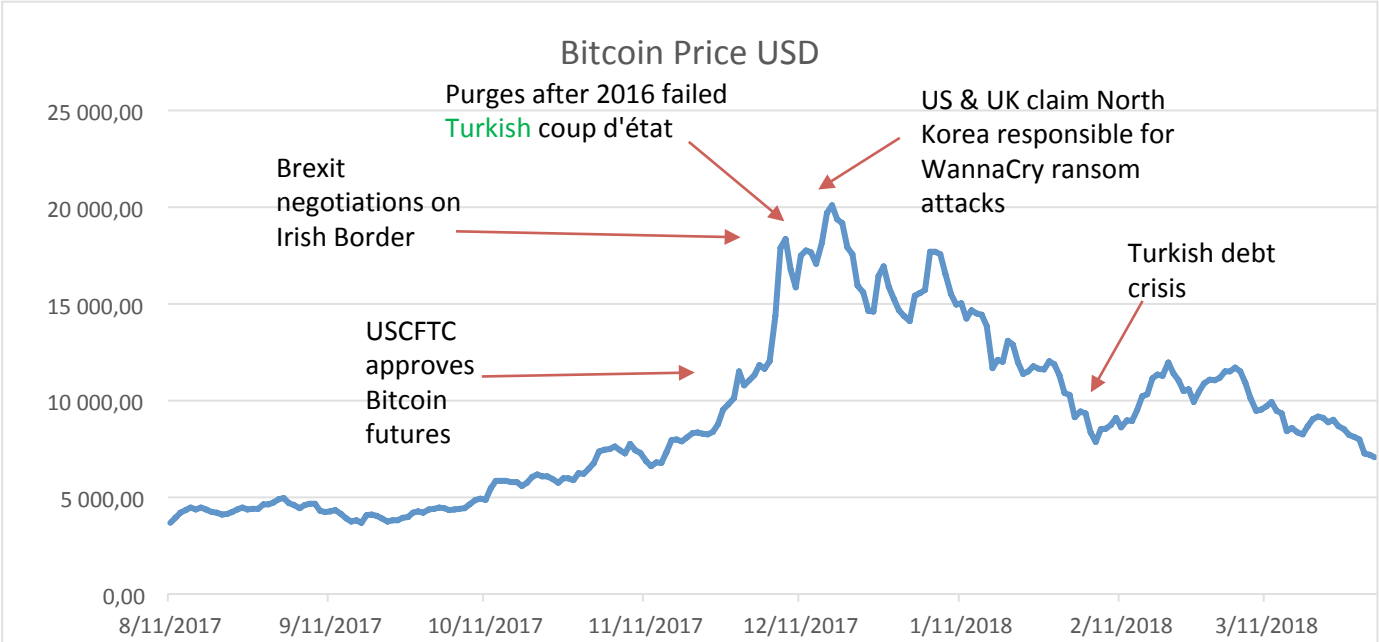

Figure 2: Log standardised prices (''daily high'', see text) of cryptocurrencies from 11/08/2017 to 1/04/2018.

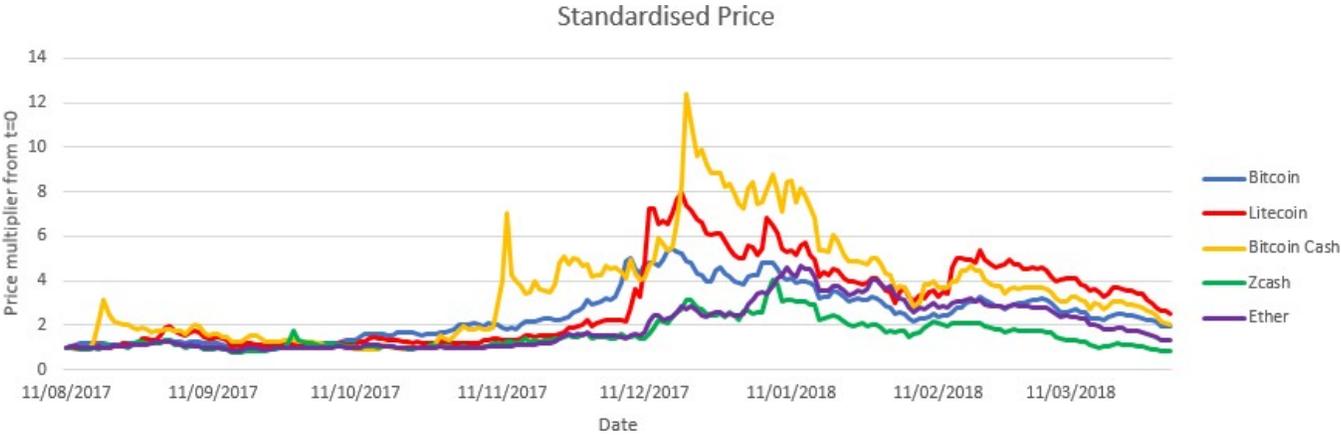

Figure 3: Most popular worldwide Google searches from April 2017-2018 by location; sourced from Trends.google.com.

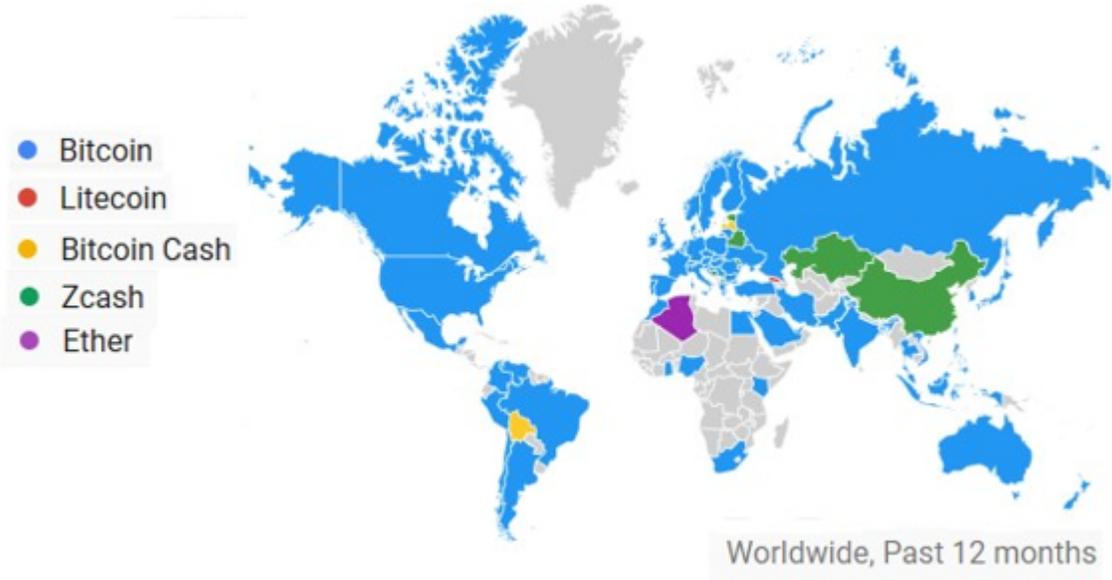

# References

Ammous, S. (2018). Can cryptocurrencies fulfil the functions of money?. The Quarterly Review of Economics and Finance, 70, 38-51.

Barber, S., Boyen, X., Shi, E., & Uzun, E. (2012). Bitter to better - how to make bitcoin a better currency. In Keromytis, A. D. (Ed.) Financial Cryptography and Data Security: 16th International Conference, FC 2012, Revised Selected Papers, Lecture Notes in Computer Science, Vol. 7397; Springer, Berlin, Germany, pp. 399-414.

Begušić, S., Kostanjčar, Z., Stanley, H. E., & Podobnik, B. (2018). Scaling properties of extreme price fluctuations in Bitcoin markets. Physica A: Statistical Mechanics and its Applications, 510, 400-406.

BitInfoCharts. (2018). Bitcoin, Litecoin, Namecoin, Dogecoin, Peercoin, Ethereum stats. [online] Available at: < https://bitinfocharts.com/>.

Blockchain.info. (2018). Bitcoin Block Explorer - Blockchain. [online] Available at: <https://blockchain.info/>.

Bogart, S. (2017). Bitcoin Vs. Bitcoin Cash: A Story Of Prioritization & Healthy Competition In Money    [online]    Available at: <https://www.forbes.com/sites/spencerbogart/2017/11/13/bitcoin-vs-bitcoin-cash-a-story-of-prioritization-a-healthy-competition-in-money/#7460f2be4bcc>.

Bohme, R., Christin, N., Edelman, B., & Moore, T. (2015). Bitcoin: Economics, Technology, and Governance. Journal of Economic Perspectives, 29(2), 213-238.

Cerqueti, R., Giacalone, M., & Mattera, R. (2020). Skewed non-Gaussian GARCH models for cryptocurrencies volatility modelling. Information Sciences. 527, 1-26

Chatterjee, J. M., Ghatak, S., Kumar, R., & Khari, M. (2018). BitCoin exclusively informational money: a valuable review from 2010 to 2017. Quality & Quantity, 52(5), 2037-2054.

Chen, A. (2016). We Need to Know Who Satoshi Nakamoto Is. [online] The New Yorker. Availableat: http://www.newyorker.com/business/currency/we-need-to-know-who-satoshi-nakamoto-is.

Chohan, U. W., (2019). The Limits to Blockchain? Scaling vs. Decentral-

ization (February 20, 2019). Discussion Paper Series: Notes on the 21st Century (CBRI). Available at SSRN: <https://ssrn.com/abstract=3338560> or <http://dx.doi.org/10.2139/ssrn.3338560>.

Coinmarketcap.com. (2018). Top 100 Cryptocurrencies by Market Capitalization. CoinMarketCap. [online] Available at: <https://coinmarketcap.com>.

Dhesi, G. & Ausloos, M. (2016). Modelling and measuring the irrational behaviour of agents in financial markets: Discovering the psychological soliton. Chaos, Solitons & Fractals, 88, 119-125.

Drożdż, S., Minati, L., Oświęcimka, P., Stanuszek, M., & Wątorek, M. (2019). Signatures of the crypto-currency market decoupling from the Forex. *Future Internet*, *11*(7), 154.

En.wikipedia.org. (2018). Ethereum. [online] Available at: <https://en.wikipedia.org/wiki/ Ethereum>.

Ethereum.org. (2018). What is Ether (ETH)?. [online] Available at: <https://www.ethereum.org/en/eth/>.

Eyal, I., Gencer, A. E., Sirer, E. G., & van Renesse, R. (2016). Bitcoin-NG: A Scalable Blockchain Protocol. In 13th USENIX Symposium on Networked Systems Design and Implementation (INSDI), pp. 45-59.

de Filippi, P. & Loveluck, B. (2016). The invisible politics of Bitcoin: governance crisis of a decentralized infrastructure. Internet Policy Review, 5(3), 1-28.

Gandal, N., & Halaburda, H. (2016). Can We Predict the Winner in a Market with Network Effects? Competition in Cryptocurrency Market. Games, 7(3), 16.

Google Trends. (2018). Google Trends. [online] Available at: <https://trends.google.com/trends/>.

Grinberg, R. (2012). Bitcoin: an innovative alternative digital currency. Hastings Science & Technology Law Journal, 4(1), 159-207.

Guo, D., Dong, J., & Wang, K. (2019). Graph structure and statistical properties of Ethereum transaction relationships. Information Sciences, 492, 58-71.

Herrera-Joancomarti, J. (2014). Research and Challenges on Bitcoin

Anonymity, Proceedings of the 9th International Workshop on Data Privacy Management. Garcia-Alfaro, J., Herrera-Joancomarti, J., Lupu, E., Posegga, J., Aldini, A., Martinelli, F., & Suri, N. (Eds.). Springer. Lecture Notes in Computer Sciences 8872, pp. 1-14.

Hruska, J. (2014). One Bitcoin group now controls 51% of total mining power, threatening entire currency's safety. ExtremeTech. [online] Available at: <https://www.extremetech.com/extreme/184427-one-bitcoin-group-now-controls-51-of-total-mining-power-threatening-entire-currencys- safety>.

Hsu, S. (2018). China Serious About Ending ICOs, Cryptocurrency Exchanges [online] Available at: <https://www.forbes.com/sites/sarahsu/2018/02/07/china-serious-about- ending-icos-cryptocurrency-exchanges/#63cd57165675>.

Kristoufek, L. (2015). What are the main drivers of the Bitcoin price? Evidence from Wavelet Coherence Analysis. PLoS ONE 10(4), 1-15.

Li, X., & Wang, C. A. (2017). The technology and economic determinants of cryptocurrency exchange rates: The case of Bitcoin. Decision Support Systems, 95, 49-60.

Li, Z., Tao, R., Su, C. and Lobonţ, O., (2018). Does Bitcoin bubble burst?. Quality & Quantity, 53(1), 91-105.

Madrigal, A. (2017). Libertarian Dream? A Site Where You Buy Drugs With Digital Dollars. [online] The Atlantic. Available at: https://www.theatlantic.com/technology/archive/2011/06/libertarian-dream- a-site-where-you-buy-drugs-with-digital-dollars/239776/

Manavi, S.A., Jafari, GR., Rouhani, S. & Ausloos, M. (2020). Demythifying the belief in cryptocurrencies decentralized aspects. A study of cryptocurrencies time cross-correlations with common currencies, commodities and financial indices, Physica A: Statistical Mechanics and its Applications, 556, 124759.

Maurer, F.K. (2016). A survey on approaches to anonymity in Bitcoin and other cryptocurrencies. Lecture Notes in Informatics, Bonn, Germany, Informatik, 2145–2150.

Nakamoto, S. (2008). Bitcoin: A peer-to-peer electronic cash system. Downloaded from academia.edu

Ong, B., Lee,T.M., Li, G. and Chuen. D.L.K. (2015). Evaluating the Potential of Alternative Cryptocurrencies. In Handbook of Digital Currency, edited by David Lee Kuo Chuen, pp. 81-135. San Diego, CA: Academic Press.

Pieters, G. and Vivanco, S. (2017). Financial regulations and price inconsistencies across Bitcoin markets. Information Economics and Policy, 39, 1-14.

Quandl.com. (2018). Quandl. [online] Available at: <https://www.quandl.com/>.

Qin, M., Su, C. W., & Tao, R. (2020). BitCoin: A new basket for eggs?. Economic Modelling Available [online] 19 February 2020 at <https://doi.org/10.1016/j.econmod.2020.02.031>.

Reid, F., & Harrigan, M. (2013). An analysis of anonymity in the bitcoin system. In "Security and privacy in social networks", Altshuler, Y., Elovici, Y., Cremers, A. B., Aharony, N., & Pentland, A. (Eds.). (pp. 197-223). Springer, New York, NY.

Rosic, A. (2017a). What is Ethereum Classic? Ethereum vs Ethereum Classic. [online] Available at: <https://blockgeeks.com/guides/what-is-ethereum-classic/>.

Rosic, A. (2017b). What is Litecoin? A Basic Beginners Guide. [online] Available at: <https://blockgeeks.com/guides/litecoin/>.

Sapuric, S., Kokkinaki, A., & Georgiou, I. (2017). In Which Distributed Ledger Do We Trust? A Comparative Analysis of Cryptocurrencies. MCIS 2017 Proceedings 21.

Shaw, A. & Hill, B.M. (2014). Laboratories of oligarchy? How the iron law extends to peer production. Journal of Communication, 64(2), 215-238.

Shen, L. (2017). Bitcoin Futures Are About To Be A Thing And It's Sending Prices Soaring. Fortune [online] Available at: <https://fortune.com/2017/12/01/bitcoin-price-cme-cftc-futures/> .

Sovbetov, Y. (2018). Factors Influencing Cryptocurrency Prices: Evidence from Bitcoin, Ethereum, Dash, Litcoin, and Monero. Journal of Economics and Financial Analysis, 2(2), 1-27.

Su, C., Qin, M., Tao, R. and Zhang, X., (2020). Is the status of gold threatened by Bitcoin?. Economic Research-Ekonomska Istraživanja, 33(1), 420-437.

Trautman, L. J. (2014). Virtual currencies; Bitcoin & What Now after Liberty Reserve, Silk Road, and Mt. Gox?. Richmond Journal of Law and

Technology, 20(4), 1-109.

Turpin, J. B. (2014). Bitcoin: The economic case for a global, virtual currency operating in an unexplored legal framework. Indiana Journal of Global Legal Studies, 21, 335-368.

Urquhart, A., (2018). What Causes the Attention of Bitcoin?. Economics Letters, 166, 40-44.

Wang, S., & Vergne, J. P. (2017). Buzz Factor or Innovation Potential: What explains cryptocurrencies' returns? PLoS ONE, 12(1), 1–17.

Wright, C. S. (2016). Turing Complete Bitcoin Script White Paper (April 10, 2016). Available at SSRN: <https://ssrn.com/abstract=3160279>.

Zcash. (2018). What is Zcash? Zcash vs. Bitcoin, Which Wins in 2018?. [online] Buybitcoinworldwide.com. Available at: <https://www.buybitcoinworldwide.com/zcash/>.